\documentclass[pra,aps,amsfonts,amsmath,superscriptaddress,twocolumn,showpacs,floatfix]{revtex4}
\usepackage{graphicx} 
\usepackage{epsfig} 
\usepackage{subfig}
\RequirePackage{amsmath,amsfonts,amssymb,amsthm}
\usepackage{rotating}
\def\be{\begin{equation}}
\def\ee{\end{equation}}
\def\bea{\begin{eqnarray}}
\def\eea{\end{eqnarray}}

\begin{document}

\title{Relativistic Hartree-Fock-Bogoliubov model for deformed nuclei }

\author{J.-P. Ebran}
\affiliation{Institut de Physique Nucl\'eaire, Universit\'e Paris-Sud, IN2P3-CNRS, F-91406 Orsay Cedex, France}
\author{E. Khan}
\affiliation{Institut de Physique Nucl\'eaire, Universit\'e Paris-Sud, IN2P3-CNRS, F-91406 Orsay Cedex, France}
\author{D. Pe\~na Arteaga}
\affiliation{Institut de Physique Nucl\'eaire, Universit\'e Paris-Sud, IN2P3-CNRS, F-91406 Orsay Cedex, France}
\author{D. Vretenar}
\affiliation{Physics Department, Faculty of Science, University of Zagreb, 10000 Zagreb, Croatia}

\begin{abstract}
The Relativistic Hartree-Fock-Bogoliubov model for axially deformed nuclei (RHFBz) is introduced. The model is based on an effective Lagrangian with density-dependent meson-nucleon couplings in the particle-hole channel, and the central part of the Gogny force is used in the pairing channel. The RHFBz quasiparticle equations are solved by expansion in the basis of a deformed harmonic oscillator. Illustrative RHFBz calculations are performed for Carbon, Neon and Magnesium isotopes. The effect of explicitly including the pion field is investigated for binding energies, deformation parameters, and charge radii and has an impact on the nuclei's shape.
\end{abstract}


\date{\today}

\maketitle

\section{Introduction}

Among the microscopic approaches to the nuclear many-body problem, nuclear energy density functionals (EDF) represent a tool of choice for the description of both static and dynamic properties of nuclei over the whole nuclide chart. EDFs subsum nucleonic short-range in-medium correlations, whereas static long-range correlations (deformation, pairing, ...) are incorporated by allowing a single-determinant state to break the symmetries of the nuclear Hamiltonian~\cite{lac09}. A variety of structure phenomena in stable and exotic nuclei have successfully been described by EDFs based on the non-relativistic Gogny and Skyrme~\cite{bend03} effective interactions, as well as on relativistic phenomenological Lagrangian densities~\cite{vret05}. 
There are significant advantages in using covariant functionals \cite{FS.00}. 
The most obvious is the natural inclusion of the
nucleon spin degree of freedom, and the resulting nuclear spin-orbit potential
which emerges automatically with the empirical strength in a covariant formulation.
The consistent treatment of large, isoscalar, Lorentz scalar and vector self-energies
provides a unique parametrization of time-odd components of the nuclear
mean-field~\cite{AA10}, i.e. nucleon currents, which is absent in the non-relativistic
representation of the energy density functional. The empirical pseudospin
symmetry in nuclear spectroscopy finds a natural explanation in terms of
relativistic mean fields \cite{Joe.05}. A covariant
treatment of nuclear matter provides a distinction between scalar and
four-vector nucleon self energies, leading to a very natural saturation
mechanism.

An example of a covariant EDF is known as the Relativistic Mean Field (RMF) framework~\cite{vret05}. The corresponding effective Lagrangians provide a quantitative description of a variety of ground-state data (masses, charge radii, ...). RMF, however, does not consider explicitly the Fock term. The exchange contributions are implicitly taken into account through the fit of model parameters to structure data. A more involved approach, the Relativistic Hartree-Fock (RHF) theory ~\cite{bouy87}, includes the exchange contributions explicitly. Early RHF models used to predict  nuclei that were considerably under-bound compared to experiment. The reason was the lack of a medium dependence in the corresponding effective nucleonic interaction~\cite{bouy87}. The effect of the nuclear medium was first taken into account by adding self-interaction terms for the $\sigma$ meson field~\cite{bern93}. Although some improvement was obtained, the RHF results were still not on the level  of RMF model predictions. An explicit nucleon-density dependence of the nucleon-meson couplings was included in Ref.~\cite{long06}. This brought a significant improvement, so that current RHF models  provide a quantitative description of nuclear properties with a similar accuracy as the standard RMF approach~\cite{long10}. In particular, recent studies by W.H. Long \textit{et. al.}~\cite{long06,long10,long08,long07} and H. Liang \textit{et. al.}~\cite{liang08,liang10} have shown that, compared to the RMF approach, the explicit treatment of Fock terms can improve the description of nuclear matter and finite nuclei.  Moreover, it explicitly takes into account the tensor contributions to the inter-nucleon interaction generated by the exchange of the $\pi$ and $\rho$ mesons. These contributions have been found to play an important role in the description of the evolution of shell structures in the framework of the shell model~\cite{ots05}.  The explicit treatment of exchange contributions in covariant EDF models enables  the inclusion of the pion field, which contributes only via its Fock term, and the tensor $\rho$-nucleon coupling, which contributes predominantly in the exchange channel. The pion contribution is expected to improve the predicted evolution of shell structure~\cite{long08}, whereas the inclusion of the tensor $\rho$ -nucleon coupling cures artificial shell gaps that arise in covariant EDF based model~\cite{long07}. 

Another benefit brought by the explicit treatment of the Fock term deals with the RPA description of collective excitations. For example, the RHF+RPA approach provides a fully self-consistent description of charge-exchange excitations~\cite{liang08}, in contrast to the RMF+RPA model in which additional 
terms have to be introduced~\cite{paar04}.

When considering applications an important challenge for the
framework of EDF is the systematic treatment of collective correlations related to
restoration of broken symmetries and fluctuations in collective coordinates.
A static nuclear EDF is characterized by symmetry
breaking -- translational, rotational, particle number, and can only
provide an approximate description of bulk ground-state properties.
To calculate excitation spectra and electromagnetic transition rates
in individual nuclei, it is necessary to extend the self-consistent mean-field 
scheme to include correlations that arise from symmetry restoration and 
fluctuations around the mean-field minimum. RMF-based models have recently 
been developed that include the explicit treatment of collective correlations, and 
employed in spectroscopic studies of a variety structure phenomena related to
shell evolution \cite{NVR.11}. 

The RHF framework has so far been limited to the description of spherical nuclei. In this 
work we consider an extension of this approach to deformed, axially-symmetric nuclei, and introduce the Relativistic Hartree-Fock-Bogoliubov model with density-dependent meson-nucleon couplings (RHFBz). 
In Sec. \ref{sec:nm} the general formalism of the RHFBz model is presented. In Sec.~\ref{Applications} we present and discuss the first applications of the RHFBz model to ground-state properties of carbon, neon and magnesium isotopes. To compare the results with those obtained with the standard RHB model \cite{vret05}, calculations are first performed without the inclusion of the pion field (PKO2 parametrization~\cite{long06a,longthesis}). The effects induced by the pion field are analyzed  in RHFBz calculations based on the PKO3 parameter set~\cite{long06a,longthesis}. Finally, Sec.~\ref{Conclusion} 
contains a short summary and discussion of possible future studies.

\section{Formalism of the RHFBz model}
\label{sec:nm}

\subsection{Energy Density Functional}

 \subsubsection{Effective Lagrangian and equations of motion}

The RHFBz approach is based on a phenomenological Lagrangian density formulated in terms of relevant degrees of freedom for nuclear structure, namely nucleons and mesons. Nucleons are treated as point-like Dirac particles. The effective in-medium interaction between nucleons is described by meson exchange, whereas the Coulomb interaction between protons is taken into account by the electromagnetic 4-potential $A^\mu$. The Lagrangian density reads :
 \begin{eqnarray}
  \left. \mathcal{L}\right.&& = \bar{\psi} \left\lbrace  i\gamma^{\mu}\partial_{\mu}-M-g_{\sigma}(\rho_v)\sigma -g_{\omega}(\rho_v)\gamma_{\mu}\omega^{\mu} \phantom{\frac{f_{\pi}(\rho_v)}{m_{\pi}}} \right. \nonumber \\ && \left. -g_{\rho}(\rho_v)\gamma_{\mu}\vec{\rho}.\vec{\tau}^{\mu}   -\frac{f_{\pi}(\rho_v)}{m_{\pi}}\gamma_5\gamma_{\mu}\partial^{\mu}\vec{\pi}.\vec{\tau} 
-e\gamma_{\mu} A^{\mu} \frac{1-\tau_3}{2}\right\rbrace  \psi \nonumber \\
&& +\frac{1}{2}\left( \partial_{\mu}\sigma\partial^{\mu}\sigma - m_{\sigma}^2 \sigma^2 \right) 
 -\frac{1}{2}  \left( \Omega_{\mu\nu}  \Omega^{\mu\nu}- m_{\omega}^2 \omega_{\mu}\omega^{\mu} \right) \nonumber \\ 
 &&-\frac{1}{2}  \left( \vec{\mathcal{R}}_{\mu\nu}\vec{ \mathcal{R}}^{\mu\nu}- m_{\rho}^2 \vec{\rho}_{\mu}\vec{\rho}^{\mu} \right) \nonumber \\
 && +\frac{1}{2}  \left( \partial_{\mu} \vec{\pi}\partial^{\mu}\vec{ \pi}- m_{\pi}^2 \vec{\pi}^2 \right) 
 -\frac{1}{2} \left(\mathcal{F}_{\mu\nu}\mathcal{F}^{\mu\nu} \right) \;.
 \label{lagrangian}
 \end{eqnarray}
Vectors in isospin space are denoted by arrows, and boldface symbols
will indicate vectors in ordinary three-dimensional space. The Dirac
spinor $\psi$ denotes the nucleon with mass $M$.  $m_\sigma$,
$m_\omega$, $m_\rho$, and $m_\pi$ are the masses of the $\sigma$-meson, the
$\omega$-meson, the $\rho$-meson and the $\pi$-meson, respectively.  
$g_\sigma$, $g_\omega$, $g_\rho$ and $f_\pi$ are the corresponding 
coupling constants for the mesons to
the nucleon. $e^2 /4 \pi = 1/137.036$. The (density-dependent) coupling constants and
meson masses are parameters, adjusted to reproduce nuclear
matter properties and ground-state properties of finite nuclei.
$\Omega ^{\mu \nu }$, $\vec{R}^{\mu \nu }$, and $\cal{F}^{\mu \nu }$ are
the field tensors of the vector fields $\omega $, $\rho $, and of
the photon \cite{bouy87}.
A nucleon-density dependence of the meson-nucleon couplings accounts for medium polarisation and three-body correlations \cite{TW.99,NVF.02,long06}. 
The effective Lagrangian is, therefore, characterized by 8 free parameters:
\begin{eqnarray}
&& m_\sigma, \  g_\sigma(\rho_{sat}),\ b_\sigma,\ d_\sigma \nonumber \\
&& g_\omega(\rho_{sat}),\ b_\omega \nonumber \\
&& \ a_\rho                                    \nonumber \\
&& \ a_\pi 
\end{eqnarray}
that are adjusted in a fit to experimental masses of twelve spherical nuclei $( ^{16}$0, $^{40}$Ca, $^{48}$Ca, $^{56}$Ni, $^{68}$Ni, $^{90}$Zr, $^{116}$Sn, $^{132}$Sn, $^{182}$Pb, $^{194}$Pb, $^{208}$Pb, $^{214}$Pb), as well as to nuclear matter properties (saturation point, 
incompressibility modulus $K_\infty$, and symmetry energy at saturation $J$).

The single-nucleon Dirac equation is derived by variation of the
Lagrangian (\ref{lagrangian}) with respect to $\bar{\psi}$
\begin{equation} \label{EqDirac1}
\left[  i\gamma^\mu \partial_\mu - M - \Sigma \right]  \psi(x) = 0 \; ,
\end{equation}
where $\Sigma $ stands for the nucleon self-energy.
When the variation is taken with respect to the boson fields, a set of inhomogeneous Klein-Gordon equations is obtained:
 \begin{eqnarray} \label{EqKG}
 && \left(\Box + m_\sigma^2 \right)\sigma = -g_\sigma \bar{\psi}\psi                                       \\
 && \left(\Box + m_\omega^2 \right)\omega^\mu = g_\omega \bar{\psi}\gamma^\mu\psi               \\
 && \left(\Box + m_\rho^2\right) \vec{\rho^\mu} = g_\rho \bar{\psi}\gamma^\mu\vec{\tau}\psi \\
 &&  \left(\Box + m_\pi^2 \right)\vec{\pi} = \frac{f_\pi}{m_\pi} \partial_\mu \left[ \bar{\psi}\gamma^5     \gamma^\mu\vec{\tau}\psi\right]                                                                           \\ 
 && \Box A^\mu  = e \bar{\psi}\gamma^\mu\frac{1-\tau_3}{2}\psi                                
\end{eqnarray}
where the conservation of the baryonic current $j^\mu = \bar{\psi}\gamma^\mu\psi $ and the Coulomb gauge choice $\left(\partial_\mu A^\mu = 0 \right)  $ have been taken into account. The Hamiltonian of the model is derived from a Legendre transformation of the Lagrangian:
\begin{eqnarray} \label{ham}
 && H = \int{d^3x \bar{\psi} \left[ -i \bm{\gamma}.\bm{\nabla}  + M \right] \psi} \nonumber \\
  &&+\frac{1}{2} \int d^3x\  \bar{\psi} \  [\ g_\sigma \sigma + g_\omega \gamma_\mu \omega^\mu + g_\rho \gamma_\mu \vec{\rho^\mu}.\vec{\tau}  \nonumber \\
&&+\frac{f_\pi}{m_\pi} \gamma_5 \bm{\gamma}.\bm{\nabla}\vec{\pi}.\vec{\tau} + e \gamma_\mu A^\mu \frac{1-\tau_3}{2} \  ]\psi \;.
\end{eqnarray}
\subsubsection{Inclusion of the Fock term}

To explicitly include the exchange contributions, it is convenient  to eliminate the mesonic degrees of freedom in Eq. (\ref{ham}) using the formal solution of the Klein-Gordon equations (Eq. (\ref{EqKG})):
\begin{eqnarray} \label{HamExact}
&& H = \int{d^3x \bar{\psi} \left[ -i \bm{\gamma}.\bm{\nabla}  + M \right] \psi} \nonumber \\ && +\frac{1}{2} \int{ d^3x d^4 y\ \bar{\psi}(x)\bar{\psi}(y) \Gamma_m(x,y) D_m(x,y)\  \psi(y) \psi(x) } \nonumber \\
\end{eqnarray} 
where a summation over the repeated index $m = \left\lbrace \sigma, \omega, \rho, \pi, A \right\rbrace $ is implied. $D_m(x,y)$ represents the propagator of the boson $m$,  whereas $\Gamma_m(x,y)$ corresponds to 2-body interaction matrices \cite{bouy87}.
The nucleon field is quantized, and the {\em no-sea} approximation ~\cite{furn04} is adopted for 
the nucleon states. The nucleon field operator $\psi$ can be expanded on an auxiliary one-body operator basis $\{c_i,c^\dagger_i\}$:
\begin{eqnarray} \label{DvlptPsinosea}
 &&  \psi(x) = \sum_i \left\lbrace  f_i(\bm{x}) e^{-i\varepsilon_i t} c_i \right\rbrace     \\
 &&  \psi^\dagger(x) = \sum_i\left\lbrace f_i^\dagger(\bm{x}) e^{i\varepsilon_i t} c_i^\dagger \right\rbrace \;.     
\end{eqnarray}
The Hamiltonian (\ref{HamExact}) consequently takes the form:
\begin{equation} \label{HamNoSea}
  H  = T + \sum_m V_m 
\end{equation}
where 
\begin{eqnarray} \label{EkinetEpot}
 && T = \sum_{i,j} c^\dagger_i c_j\int{d^3x \bar{f_i} \left[ -i \bm{\gamma}.\bm{\nabla}  + M \right] f_j} \nonumber \\
 && V_m = \frac{1}{2} \sum_{i,j,k,l}c^\dagger_i c^\dagger_j c_kc_l \int d^3x_1 d^3x_2\ \bar{f_i}(\bm{x_1})\bar{f_j}(\bm{x_2}) \nonumber \\
&& \Gamma_m(1,2) D_m(\bm{x_1},\bm{x_2})\  f_k(\bm{x_2}) f_l(\bm{x_1})  \;.
\end{eqnarray}
The effective inter-nucleon interaction contained in the Hamiltonian (\ref{HamNoSea}) is designed to be used in the self-consistent mean-field approximation. The ground-state of the nuclear many-body system  is, therefore, approximated by a Slater determinant: 
\begin{equation} \label{EtatHF}
\left.  \mid \Phi_0 \right\rangle  =  \prod_i c^\dagger_i \left| 0 \right\rangle \;, 
\end{equation}
where $\left. \mid 0 \right\rangle $ represents the single-nucleon vacuum.
The energy density functional is then obtained by taking the expectation value of the Hamiltonian (\ref{HamNoSea}) in the ground-state Slater determinant (\ref{EtatHF}):
 \begin{equation}
 \left. \mathcal{E}\right.^{RHF}\left[\rho \right] = \left\langle \Phi_0\mid H\left[\rho \right]  \mid \Phi_0 \right \rangle \; .
\end{equation}
In particular, the expectation value of the potential energy operator generates a direct and an exchange term. This RHF functional can be written in terms of the one-body density operator represented by the matrix elements $\rho_{ij} =  \left\langle \Phi_0\mid c^\dagger_j c_i  \mid \Phi_0 \right\rangle $:
\begin{eqnarray} \label{EDF}
 \left. \mathcal{E}\right.^{RHF}\left[\rho,\phi_m \right] = Tr\left[\left(-i \bm{\gamma}.\bm{\nabla}  + M + \Gamma_m \phi_m \right) \rho  \right] \nonumber \\
+ \frac{1}{2} Tr\left[\left.\mathcal{V}\right.[\rho] \  \rho^\dagger \otimes \rho \right]  \pm \frac{1}{2}\int{d^3x\left[\left(\partial_\mu \phi_m \right)^2 + m_m^2  \right] }
\end{eqnarray}
Here $\Gamma_m$ represents the one-body vertex function \cite{bouy87}.
The trace operator involves a summation over space-time coordinates and Dirac indices. $\left.\mathcal{V} \right. $ is defined by its matrix elements $V_{ijkl}$:
\begin{eqnarray} \label{KinetPot}
 V_{ijkl} = \int d^3x_1 d^3x_2\ &&\bar{f_i}(\bm{x_1})\bar{f_j}(\bm{x_2}) \Gamma_m(1,2) D_m(\bm{x_1},\bm{x_2})\ \nonumber \\ && f_k(\bm{x_2}) f_l(\bm{x_1})  
\end{eqnarray}
Finally, the tensor product corresponds to:
\begin{equation}
\left(  \rho^\dagger \otimes \rho\right)_{ijkl} = \rho^\dagger_{ik} \  \rho_{lj}
\end{equation}

\subsection{RHF equations for systems with axial symmetry}

The minimization of the energy functional (\ref{EDF}) with the constraint that the single-nucleon density matrix refers to a Slater determinant leads to the RHF equations:
\begin{eqnarray} \label{EqRho}
 && \left[h[\rho], \rho \right] = 0  \nonumber \\
 && \left(-\bigtriangleup + m_m^2 \right) \phi_m = \pm Tr\left(\Gamma_m \rho \right) \;,  
\end{eqnarray}
where 
\begin{equation} \label{h1corps}
 h\left[\rho \right] = \frac{\delta\left. \mathcal{E}\right.^{RHF}\left[\rho \right]  }{\delta \rho} \; .
\end{equation}
In coordinate space the set of equations (\ref{EqRho}) can be written as:
\begin{eqnarray} \label{EqHF}
 &&\left\lbrace -i\bm{\alpha}.\bm{\nabla}+\beta M^*(\bm{r})+\left[ V(\bm{r})+\Sigma^R(\bm{r})\right]\right\rbrace f_i(\bm{r},q_i)
 + \left.\mathcal{F}\right._i(\bm{r}) \nonumber \\ && =\epsilon_i f_i(\bm{r},q_i)  
\end{eqnarray}
\begin{eqnarray} \label{EqHF2}
&& \left\lbrace \begin{array}{c}  
 \left(-\bigtriangleup + m_\sigma^2 \right)\sigma(\bm{r}) = -g_\sigma(\rho_v)\ \rho_s(\bm{r}) \\
\left(-\bigtriangleup + m_\omega^2 \right)\omega^0(\bm{r}) = g_\omega(\rho_v)\ \rho_v(\bm{r}) \\
\left(-\bigtriangleup + m_\rho^2 \right)\rho_3^0(\bm{r}) = g_\rho(\rho_v)\ \rho_{tv}(\bm{r}) \\
-\bigtriangleup A^0(\bm{r}) = e \ \rho_c(\bm{r})
                \end{array} \right. 
\end{eqnarray}
where $q$ denotes isospin projection quantum number, and $\bm{\alpha} \equiv \gamma^0\bm{\gamma},\ \beta\equiv\gamma^0$ are Dirac matrices. In the Dirac equation (\ref{EqHF}):
\begin{itemize}
\item[$\bullet$] $M^*=M+S(\bm{r})$ is the Dirac effective mass.
\newline
\item[$\bullet$] $S(\bm{r})$ and $V(\bm{r})$ are the Hartree terms, i.e. they represent the direct contribution to the nucleon self-energy:
\begin{eqnarray}
 && S\left(\bm{r} \right) = g_\sigma(\rho_v) \ \sigma\left(\bm{r} \right) \\
 && V\left(\bm{r} \right) = g_\omega(\rho_v) \ \omega^0\left(\bm{r} \right) + g_\rho(\rho_v) \ \rho^0_3\left(\bm{r} \right) \tau_3 + e A^0\left(\bm{r} \right) \nonumber \\ 
\end{eqnarray}
\newline
\item[$\bullet$] $\Sigma^R(\bm{r})$ denotes the rearrangement contribution. It can be divided into a direct term $\Sigma^R_H(\bm{r})$ and an exchange term $\Sigma^R_F(\bm{r})$. The direct contribution to the rearrangement term reads:
 \begin{eqnarray}
 \Sigma^R_H(\bm{r}) = &&\frac{\partial g_\sigma}{\partial \rho_v}\rho_s(\bm{r}) \sigma(\bm{r}) + \frac{\partial g_\omega}{\partial \rho_v}\rho_v(\bm{r}) \omega_0(\bm{r}) \nonumber \\ &&+ \frac{\partial g_\rho}{\partial \rho_v} \rho_{tv}(\bm{r})\rho_0^3(\bm{r})
\end{eqnarray}
Taking the $\sigma$ meson as example, the exchange contribution to the rearrangement term reads:\begin{eqnarray}
\Sigma^{R,\sigma}_F(\bm{r}) = &&\sum_{k,l}\delta_{q_k,q_l} [\frac{\partial g_\sigma}{\partial \rho_v}\bar{f_k}(q_k)f_l(q_l) ](\bm{r}) \nonumber \\ &&\int {d{\bf
r'} \left\lbrace D_\sigma(\bm{r},\bm{r'})[g_\sigma \bar{f_l}(q_l)f_k(q_k)](\bm{r'}) \right\rbrace} 
\end{eqnarray}
\newline
\item[$\bullet$] $\left.\mathcal{F}\right._i(\bm{r})$ denotes to the Fock terms, i.e. the exchange contribution to the nucleon self-energy. For instance, the Fock term associated to the $\sigma$ 
meson:
\begin{eqnarray}
\left.\mathcal{F}\right._i^\sigma(\bm{r}) =&& \sum_{j}\delta_{q_j,q_i}\int d{\bf
r'}\left\lbrace\right.  D_\sigma(\bm{r},\bm{r'})\nonumber \\ &&\left. [g_\sigma \bar{f_j(q_j)}f_i(q_i)](\bm{r'}) \right\rbrace \beta [g_\sigma f_j](\bm{r},q_j) \nonumber \\
\end{eqnarray}
\newline
\end{itemize}
The sources of the inhomogeneous Klein-Gordon equations (\ref{EqHF2}) read :
\begin{eqnarray}
 && \rho_s(\bm{r}) = \sum_i \bar{f_i}(\bm{r}) f_i(\bm{r}) \\
 && \rho_v(\bm{r}) = \sum_i f_i^\dagger(\bm{r}) f_i(\bm{r}) \\
 && \rho_{tv}(\bm{r}) =\sum_i f_i^\dagger(\bm{r})\tau_3 f_i(\bm{r}) = \rho_v^{proton}(\bm{r}) - \rho_v^{neutron}(\bm{r}) \nonumber \\
\\
 && \rho_c(\bm{r}) = \sum_i f_i^\dagger(\bm{r}) \frac{1-\tau_3}{2} f_i(\bm{r}) = \rho_v^{proton}(\bm{r}) \\
\end{eqnarray}
In the case of deformed nuclei characterized by axial symmetry, the label $i$ of the single-nucleon wave function $f_i(\bm{r})$ refers to the set of quantum numbers:
\begin{equation}
 i=(\Omega,\Pi,q) \;,
\end{equation}
where $\Omega$ denotes the projection of the total angular momentum on the symmetry axis, $\Pi$ is the parity, and $q$ the isospin projection that distinguishes protons and neutrons. In cylindrical coordinates $(r_\perp,\phi,z)$ the nucleon wave function takes the form~\cite{ring97} :
\begin{equation}\label{eq:cyl1}
f_i(\bm{r})=\frac {1}{\sqrt{2\pi}} \left( \begin{array}{c} F_i^+(r_\perp,z;q_i) e^{i(\Omega_i-\frac{1}{2})\varphi} \\
F_i^-(r_\perp,z;q_i)e^{i(\Omega_i+\frac{1}{2})\varphi} \\
iG_i^+(r_\perp,z;q_i)e^{i(\Omega_i-\frac{1}{2})\varphi} \\
iG_i^-(r_\perp,z;q_i)e^{i(\Omega_i+\frac{1}{2})\varphi}  \end{array} \right) 
\end{equation}
The RHF equations are solved by expanding the nucleon spinors and meson fields in the basis of a deformed harmonic oscillator. The eigenfunctions $\Phi_\alpha(\bm{r}), \alpha=\left\lbrace n_z,n_r,m_l,m_s\right\rbrace $ of the deformed harmonic oscillator potential
\begin{equation}\label{OHpot}
V_{osc}(r_\perp,z) = \frac{1}{2} M \omega_\perp^2 r_\perp^2 + \frac{1}{2} M \omega_z^2 z^2 
\end{equation}
are expressed in terms of Laguerre and Hermite polynomials. They form a basis on which the nucleon wave functions $f_i(\bm{r})$ are expanded~\cite{ring97}: 
\begin{eqnarray}
 && F_i(\bm{r},q_i) = \frac {1}{\sqrt{2\pi}} \left( \begin{array}{c}
 F_i^+(r_\perp,z;q_i) e^{i(\Omega_i-\frac{1}{2})\varphi} \\
F_i^-(r_\perp,z;q_i)e^{i(\Omega_i+\frac{1}{2})\varphi} \end{array}\right) \nonumber \\ && = \sum_{\alpha}f_\alpha^{(i)}(q_i)\phi_\alpha(r_\perp,z,\varphi)    \nonumber \\
\\
 && G_i(\bm{r},q_i) = \frac {1}{\sqrt{2\pi}} \left( \begin{array}{c}
 G_i^+(r_\perp,z;q_i) e^{i(\Omega_i-\frac{1}{2})\varphi} \\
G_i^-(r_\perp,z;q_i)e^{i(\Omega_i+\frac{1}{2})\varphi} \end{array}\right) \nonumber \\ &&  = \sum_{\tilde{\alpha}}g_{\tilde{\alpha}}^{(i)}(q_i)\phi_{\tilde{\alpha}}(r_\perp,z,\varphi) \; .\nonumber \\ 
\end{eqnarray}
In a deformed harmonic oscillator basis, therefore, the solution of the Dirac equation (\ref{EqHF}) 
corresponds to a diagonalization of the matrix:
\begin{equation}\label{eq:mat1}
 \left( \begin{array} {cc}
 A_{\alpha,\alpha'} 
 & B_{\alpha,\tilde{\alpha'}} \\
 B_{\tilde{\alpha},\alpha'} 
 & C_{\tilde{\alpha},\tilde{\alpha'}} \end{array} \right)  \left( \begin{array} {c}
 f_{\alpha'}^{(i)}(q_i) \\
 g_{\tilde{\alpha'}}^{(i)}(q_i)\end{array} \right)=\varepsilon_i \left( \begin{array} {c}
 f_{\alpha}^{(i)}(q_i) \\
 g_{\tilde{\alpha}}^{(i)}(q_i)\end{array} \right)
\end{equation}
Detailed expressions for the Fock contribution to the matrices $A$, $B$, and $C$ are given in appendix~\ref{app}. 

\subsection{Pairing correlations}

For a quantitative analysis of open-shell nuclei, both spherical and deformed, it is necessary to
consider also pairing correlations. The nucleonic pairing is treated in the context of the Bogoliubov framework~\cite{rs80}. The resulting RHFB model provides a unified description of particle-hole $(ph)$ and particle-particle $(pp)$ correlations on a mean-field level by using two average potentials: the self-consistent mean field that encloses all the long range \textit{ph} correlations, and a pairing field $\hat{\Delta}$ which sums up the \textit{pp}-correlations. Pairing correlations in nuclei are restricted to an energy window of a few MeV around the Fermi level, and their scale is well separated from the scale of binding energies, which are in the range of several hundred to thousand MeV. There is no empirical evidence for any relativistic effect in the nuclear pairing field $\hat{\Delta}$ and, therefore, a hybrid RHFB model with a non-relativistic pairing interaction can be formulated. Similar to most applications of the RHB model~\cite{vret05}, the central part of the Gogny force~\cite{BGG.84} will be employed in the particle-particle ($pp$) channel:
\begin{eqnarray}
V^{pp}(1,2)~&=&~\sum_{i=1,2}e^{-((\mathbf{r}_{1}-\mathbf{r}_{2})/{\mu_{i}})^{2}%
}\,(W_{i}~+~B_{i}P^{\sigma} \nonumber \\
&&-H_{i}P^{\tau}-M_{i}P^{\sigma}P^{\tau})\;,
\label{Gogny}%
\end{eqnarray}
with the set D1S~\cite{BGG.91} for the parameters $\mu_{i}$, $W_{i}$, $B_{i}$, $H_{i}$, and $M_{i}$ $(i=1,2)$. A basic advantage of the Gogny force is the finite range, which automatically guarantees a proper cut-off in momentum space.
\section{Results and discussion}
\label{Applications}

\begin{figure}[htb]
\begin{center}
\scalebox{0.35}{\includegraphics{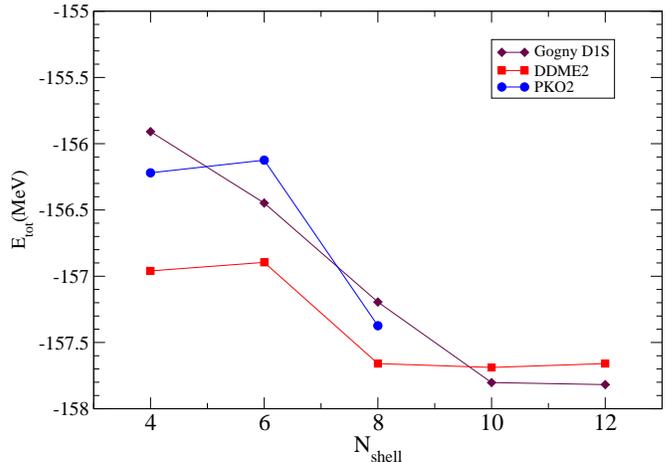}}
\caption{(Color online) Evolution of the calculated binding energy of the $^{20}$Ne nucleus with the number of major shell. The theoretical values are obtained with the PKO2~\cite{long06a,longthesis}, DDME2~\cite{vret05} and Gogny D1S~\cite{webGogny} effective interactions.}
\label{fig:ENshell}
\end{center}
\end{figure}

The explicit treatment of exchange contributions requires the calculation of non-separable two-dimensional integrals $I_{\alpha\beta\beta'\alpha'}$ in momentum space (cf. appendix~\ref{app}). These integrals involve a boson propagator and the functions $Q_{\alpha,\beta}$ (Eq. (\ref{fncQ})). The $Q$ functions are non-separable two-dimensional spatial integrals. Their numerical evaluation imposes considerable constraints on the size of the basis of a deformed harmonic oscillator. We have verified that a RHFBz calculation with 6 fermionic shells yields reliable results for ground-state properties of nuclei up to $Z=30$ (zinc isotopic chain). Various ground-state quantities (mass, axial deformation parameter, charge radius, chemical potential, single-particle energies,...) obtained from an expansion of the nucleon wave functions in a basis of 6 oscillator shells, agree within 1\% from those calculated with an expansion in 8 shells. Fig~\ref{fig:ENshell} displays the binding energy of the deformed $^{20}$Ne nucleus with respect to the number of fermionic shells calculated with the PKO2, DDME2 and Gogny D1S effective interactions. The $N_{shell} = 4$ calculation is unphysical but emphasizes the similar qualitative evolution of both PKO2 and DDME2 binding energies with the number of shell up to $N_{shell} = 8$. We estimate a 1\% numerical error for the PKO2 observables obtained from a 6 shells calculation, as seen on Fig~\ref{fig:ENshell}. We have also compared the results for $\left.\right.^{16}$O and $\left.\right.^{40}$Ca calculated with the expansion in 6 oscillator shells to those obtained by solving the spherical RHF equations in coordinate space, discretized on a mesh of $R_{max} = 20fm$ and with a step $a=0.1 fm$~\cite{long07}. The corresponding ground-state quantities and single-particle energies display relative variations of less than 1\%, validating the choice of 6 fermionic shells for model calculation of light nuclei. The results shown hereafter 
correspond to calculations with 6 fermionic shells for the RHFB model and 12 fermionic shells for the RHB model.An expansion in 20 oscillator shells is used for the solution of the Klein-Gordon equations in the mesonic sector for both RHFB and RHB models.

\subsection{Ground-state observables}

This section presents results of the first application of the RHFBz model in the calculation of ground-state properties of carbon, neon and magnesium nuclei. Masses, radii and shapes are fundamental characteristics of nuclei and their description presents a basic test of any model and effective force. Therefore, the RHFBz model with the PKO2 and PKO3 effective interactions~\cite{long06a,longthesis} in the particle-hole channel, and the central part of the Gogny D1S force~\cite{BGG.91} in the particle-particle channel, are used to calculate densities, masses, two-neutron drip-lines, deformations and charge radii of the $Z=6,10,12$ isotopic chains. The PKO2 parametrization corresponds to a covariant EDF that does not include explicitly the pion field. Thus, the effects of the one-pion exchange is taken into account implicitly through the fit of the model parameters to data. PKO3 parameterizes a covariant EDF that explicitly includes the pion degree of freedom. Neither PKO2 nor PKO3 include the tensor $\rho$-nucleon coupling. The values of the PKO2 and PKO3 parameters are listed in Table~\ref{tabpko2}, together with those of 
one of the most successful RMF functionals: DD-ME2~\cite{lal05}, that has extensively been used in applications of the RHB model. In this section, to make a first study of the influence of the Fock term on a similar ground, we compare RHFBz results obtained without the inclusion of the pion field (PKO2 effective interaction) with those of the axial RHB model (DD-ME2 effective interaction). 
 \begin{table}[h]
 \centering
\begin{tabular}{c|c|c|c}
   \phantom{PKO2}             &  PKO2          &  PKO3              & DD-ME2              \\
 \hline
  $m_\sigma$ (MeV)            &  534.461792    &  525.667664        & 555.1238           \\
  $m_\omega$ (MeV)            &  783.000000    &  783.000000        & 783.0000           \\
  $m_\rho$ (MeV)              &  769.000000    &  769.000000        & 769.0000           \\
  $g_\sigma(\rho_{sat})$      &  8.920597      &  8.895635          & 10.5396            \\ 
  $g_\omega(\rho_{sat})$      &  10.550553     &  10.802690         & 13.0189            \\ 
  $g_\rho(\rho_{sat})$        &  2.163268      &  2.030285          & 3.6836             \\
  $f_\pi(0)$                  &  0.000000      &  1.000000          & 0.0000             \\
  $a_\sigma$                  &  1.375772      &  1.244635          & 1.3881             \\ 
  $b_\sigma$                  &  2.064391      &  1.566659          & 1.0943             \\ 
  $c_\sigma$                  &  3.052417      &  2.074581          & 1.7057             \\ 
  $d_\sigma$                  &  0.330459      &  0.400843          & 0.4421             \\
  $a_\omega$                  &  1.451420      &  1.245714          & 1.3892             \\ 
  $b_\omega$                  &  3.574373      &  1.645754          & 0.9240             \\ 
  $c_\omega$                  &  5.478373      &  2.177077          & 1.4620             \\ 
  $d_\omega$                  &  0.246668      &  0.391293          & 0.4775             \\
  $a_\rho$                    &  0.631605      &  0.635336          & 0.5647             \\ 
  $a_\pi$                     &  0.000000      &  0.934122          & 0.0000             \\ 	 
\hline
\end{tabular}
\caption{\label{tabpko2} Parameters of the PKO2, PKO3~\cite{long06a,longthesis} and DD-ME2~\cite{lal05} effective interactions. }
\end{table}

\begin{sidewaysfigure*}
\centering
\scalebox{0.12}
{\includegraphics{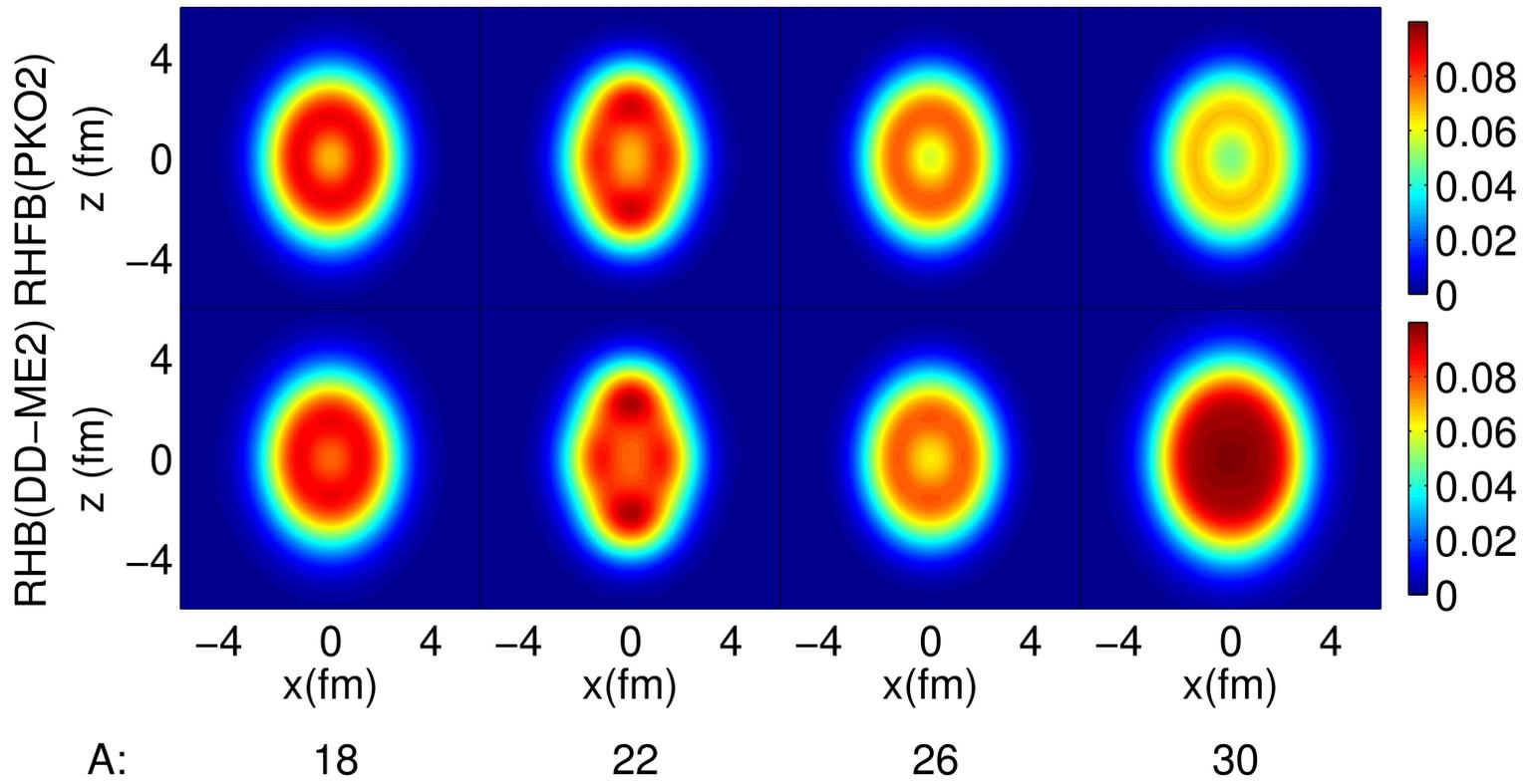}}
\caption{(Color online) Proton density in the Ne isotopic chain. The single-nucleon densities calculated with the PKO2 (RHFBz) and DD-ME2 (RHB) 
effective interactions are plotted in the (Oxz) plan with $x,z\in \left[-6fm,6fm \right]$. 
The color code 
denotes densities in the interval $\left[0fm^{-3},0.09fm^{-3} \right]$ }
\label{figZ-10_dens_bis}
\end{sidewaysfigure*}

\begin{sidewaysfigure*}
\centering
\scalebox{0.12}
{\includegraphics{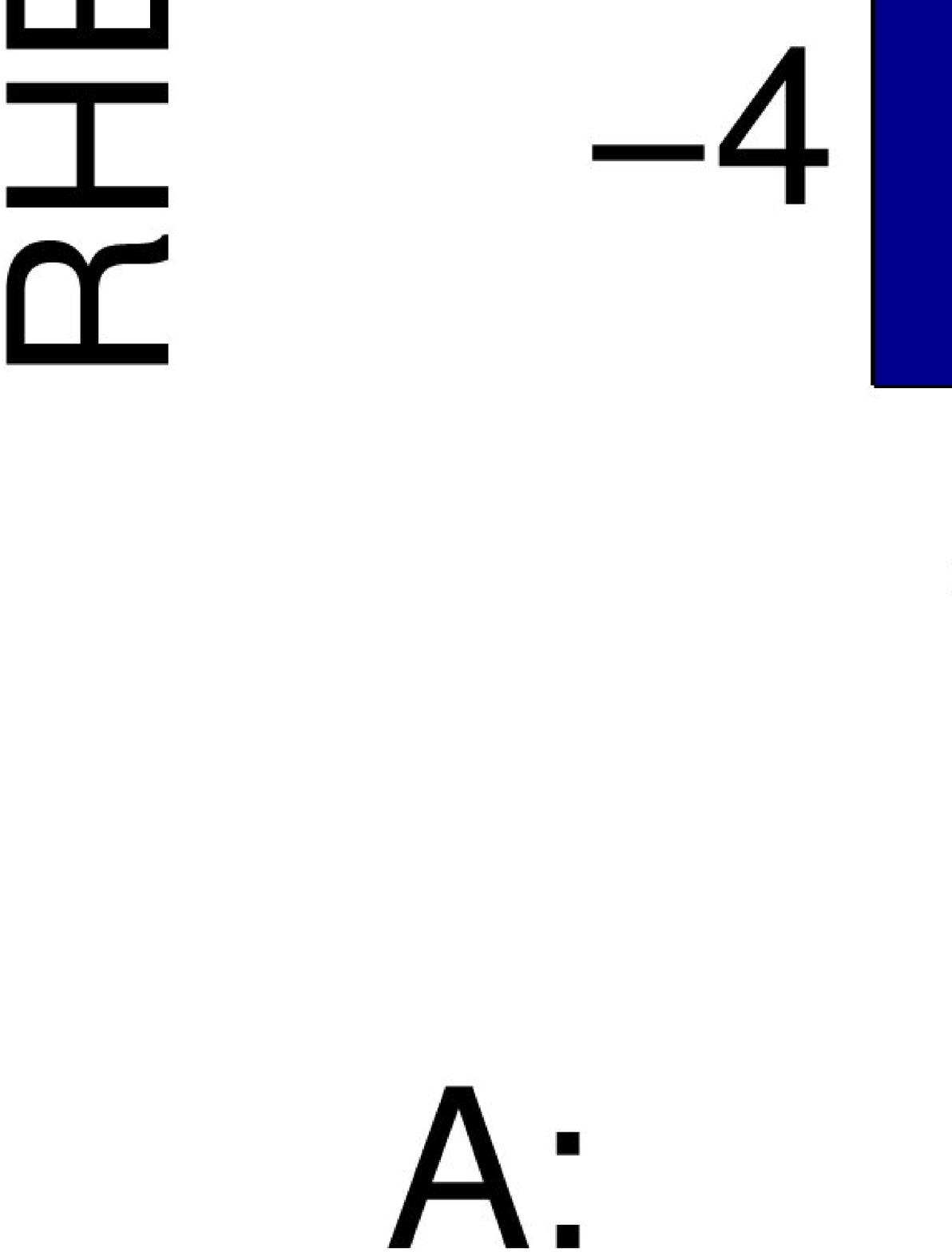}}
\caption{(Color online) Same as Fig.~\ref{figZ-10_dens_bis} for neutron density in the interval $\left[0fm^{-3},0.1fm^{-3} \right]$. }
\label{figZ-10_dens_bis2}
\end{sidewaysfigure*}

Fig.~\ref{figZ-10_dens_bis} and Fig.~\ref{figZ-10_dens_bis2} compare the proton and neutron densities of neon isotopes calculated with the PKO2 and DD-ME2 effective interactions. It appears that both models 
predict rather similar shapes for nuclei with $A \leq 26$, whereas for heavier Ne isotopes larger deformations are calculated with the RHB model with DD-ME2, especially 
for proton densities.

\begin{figure}[htb]
\begin{center}
\scalebox{0.35}{\includegraphics{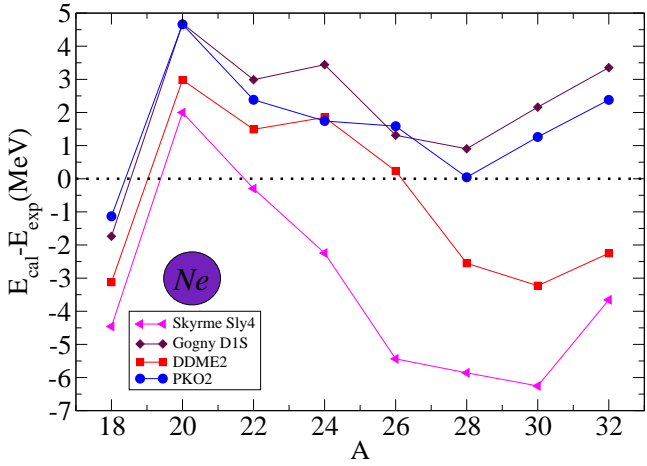}}
\caption{(Color online) Absolute deviations of the calculated binding energies from the experimental values of the Ne isotopic chain. The theoretical values are obtained with the PKO2~\cite{long06a,longthesis}, DDME2~\cite{vret05}, Gogny D1S~\cite{webGogny} and Skyrme SLy4~\cite{webSkyrme} effective interactions. The data are from Ref.~\cite{audi93}. For  $\left.\right.^{32}$Ne the calculated values are compared to the extrapolated binding energy. }
\label{fig:Neediff}
\end{center}
 \end{figure}

In Fig.~\ref{fig:Neediff} we display the absolute deviations of the calculated binding energies from the experimental values of neon isotopes for the two relativistic effective interactions PKO2, DD-ME2, the Gogny force D1S, and the Skyrme interaction SLy4. 
Positive deviations correspond to underbound nuclei. One might notice that deformed 
RHFB calculations with PKO2 predict binding energies with a level of agreement with data comparable to that of the Gogny D1S force and, for heavier isotopes, slightly better than RHB with DD-ME2. Much larger deviations from data are calculated with the Skyrme force SLy4. Similar results are found in the carbon and magnesium isotopic chains. This is quantified in Table~\ref{tabrms}, where we compare the root-mean-square deviations of theoretical binding energies for the $Z=6,10,12$ isotopic chains. PKO2 predictions are closest to the experimental values in the neon and magnesium isotopic chains, whereas DD-ME2 gives the smallest {\em rms} deviation for carbon nuclei. 


\begin{table}[h]
 \centering
\begin{tabular}{c|c|c|c|c}
   \phantom{PKO2}             &  PKO2          &  DDME2            & Gogny D1S     & Skyrme SLy4        \\
 \hline
  $\sigma_C$ (MeV)            &  2.144         &  1.443            & 3.185         & 2.874              \\
 \hline
  $\sigma_{Ne}$ (MeV)         &  2.263         &  2.429            & 2.750         & 4.342              \\
\hline
  $\sigma_{Mg}$ (MeV)         &  2.480         &  2.582            & 3.337         & 3.269              \\
\hline
\end{tabular}
\caption{\label{tabrms} Root-mean-square deviations from experimental data of binding energies calculated with 
the PKO2~\cite{long06a,longthesis}, DD-ME2~\cite{vret05}, Gogny D1S~\cite{webGogny} and Skyrme SLy4~\cite{webSkyrme} effective interactions from  for the carbon, neon and magnesium isotopic chains. }
\end{table}

\begin{figure}[htb]
\begin{center}
\scalebox{0.35}{\includegraphics{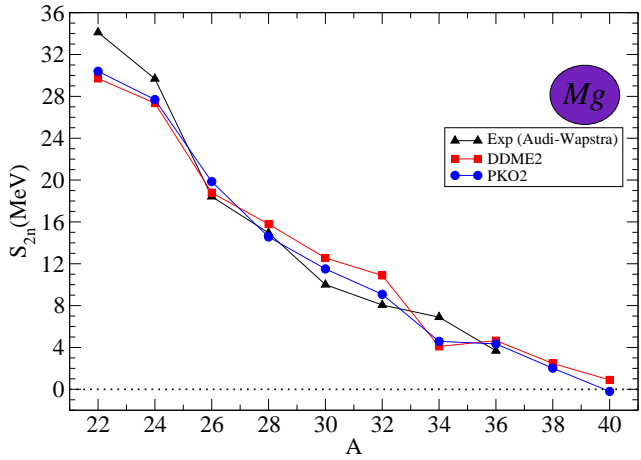}}
\caption{(Color online) Two-neutron separation energy in the magnesium isotopic chain. The relativistic mean-field results: RHFBz with PKO2~\cite{long06a,longthesis}, and RHB with DD-ME2~\cite{lal05}, are compared to data (Audi-Wapstra~\cite{audi93}).}
\label{fig:Mgs2n}
\end{center}
\end{figure}

The two-neutron separation energy $S_{2n} \equiv E_{tot}(Z,N) - E_{tot}(Z,N-2)$ of Mg 
isotopes, calculated with PKO2 and DD-ME2, are compared to data in 
Fig.~\ref{fig:Mgs2n}. In general, the RHFBz results obtained with the PKO2 parameter set are closer to the experimental two-neutron separation energies. The last two-neutron bound Mg nucleus is $\left.\right.^{38}$Mg in the RHFBz calculation with PKO2, whereas the RHB model with DD-ME2 predicts $\left.\right.^{40}$Mg to be the last bound isotope. Compared to DD-ME2, the PKO2 two-neutron separation energies are also found closer to data in the carbon and neon isotopic chains.

\begin{figure}[htb]
\begin{center}
\scalebox{0.35}{\includegraphics{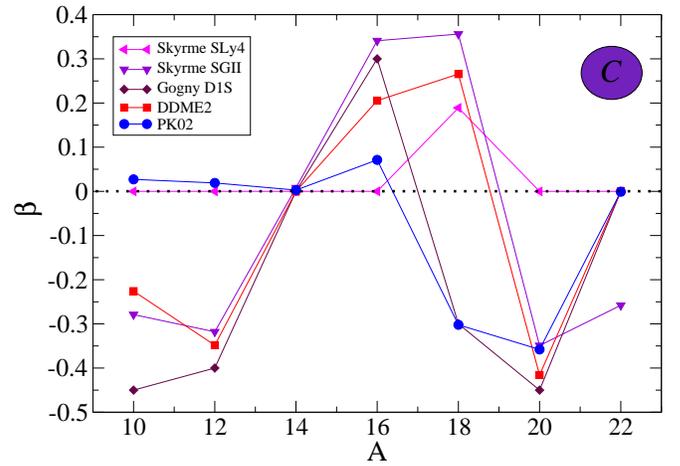}}
\caption{(Color online) Axial deformation parameter $\beta$ of C nuclei as function of the mass number. The calculated values correspond to the PKO2~\cite{long06a,longthesis}, DD-ME2~\cite{lal05}, Gogny D1S, Skyrme SLy4 and Skyrme SGII~\cite{sag04} effective interactions.}
\label{fig:Cdef}
\end{center}
 \end{figure}

The evolution of the axial deformation parameter $\beta$ with mass number along the carbon isotopic chain is displayed in Fig.~\ref{fig:Cdef}. No experimental results extracted from B(E2) measurements are presented insofar as it is not adequate to directly compare the static deformation parameter from the dynamical one extracted from the experiment, in such light even-even nuclei. PKO2 and Skyrme SLy4 predict deformations that are systematically smaller than those obtained with the DD-ME2 and Gogny D1S interactions, or with the Skyrme SGII effective force. In particular, PKO2 and 
SLy4 predict basically spherical shapes between   $\left.\right.^{10}$C and $\left.\right.^{16}$C whereas, except for  $\left.\right.^{14}$C, rather large ground-state deformations are calculated with the other three interactions. The case of $\left.\right.^{16}$C is particularly interesting. Early experiments at Riken indicated an anomalously small B(E2)~\cite{ima04}, and a strong prolate deformation~\cite{ong06}. Therefore, the nucleus $\left.\right.^{16}$C was thought to be characterized by valence neutrons decoupled from a quasi-spherical core. Different models corroborated these results (see for instance~\cite{hag07}). Recent measurements of $B(E2;2_1^+\rightarrow0^+)$  in $\left.\right.^{16}$C gave a value that is more consistent with what is observed in nuclei with similar ${N}/{Z}$~\cite{wied08,wuo10}. The current result does not support the description of $\left.\right.^{16}$C in terms of valence neutrons decoupled from the spherical core. The RHFBz model calculation with the PKO2 interaction, in particular, predicts the neutron and proton axial deformations in $\left.\right.^{16}$C: 
$\beta_n=0.08$ and $\beta_p=0.06$, respectively.
 
\begin{figure}[htb]
\begin{center}
\scalebox{0.35}{\includegraphics{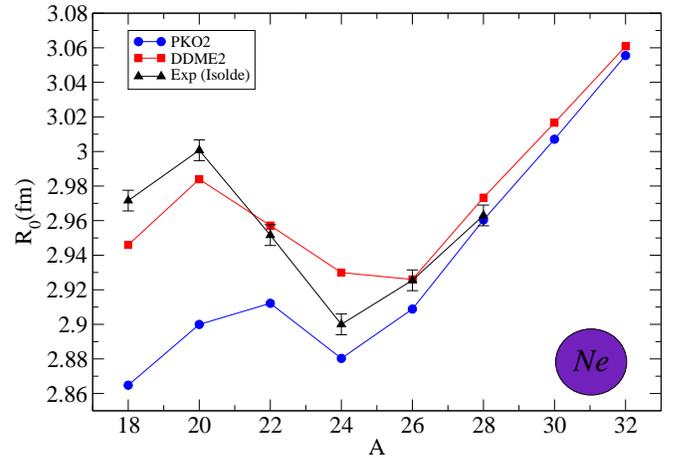}}
\caption{(Color online) Charge radii in the neon isotopic chain. The theoretical values calculated with PKO2~\cite{long06a,longthesis} and DD-ME2~\cite{lal05} are compared to the ISOLDE experimental values~\cite{marin}.The error bars do not include the atomic factor and therefore should be increased by 10\%.}
\label{fig:Nerc1}
\end{center}
 \end{figure}

The charge radii of neon isotopes, calculated with PKO2 and DD-ME2, are shown in comparison with data in Fig.~\ref{fig:Nerc1}. In lighter Ne nuclei the theoretical values predicted by the RHB model with DD-ME2 are in much better agreement with experiment, whereas for $A \geq 26$ both models yield similar charge radii. It should be noted that contrary to DD-ME2, PKO2 does not include the charge radii in its fit to spherical nuclei.

\begin{figure}[htb]
\begin{center}
\scalebox{0.35}{\includegraphics{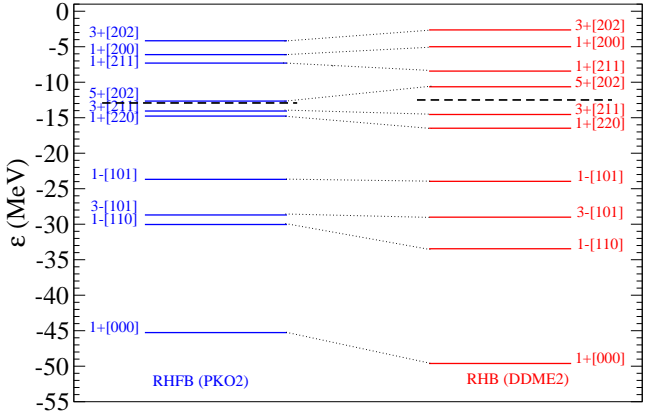}}
\caption{(Color online) Comparison of the single-proton levels of $\left.\right.^{28}$Mg, calculated with the PKO2~\cite{long06a,longthesis} and DD-ME2~\cite{lal05} effective interactions. The levels are labeled by Nilsson quantum numbers. The dashed-line denotes the chemical potential.}
\label{fig:Mgspp}
\end{center}
\end{figure}

Structure models can also be compared by considering the corresponding
single-nucleon spectra. In Ref.~\cite{longthesis} the predictions of
the spherical RHF model were tested in comparison to data for $^{16}$O
and $^{40}$Ca. In Fig.~\ref{fig:Mgspp} we compare the proton Nilsson
orbitals of $\left.\right.^{28}$Mg, calculated with PKO2 and DD-ME2. 
Although the ordering of Nilsson states is the same for both
interactions, i.e. the deformation is rather small, in general the 
density of states around the Fermi level is larger when calculated
with the RHFBz approach. This originates from the larger effective
nucleon mass characterizing the PKO2 interaction compared to the DDME2
one. Namely, the density of states depends on the effective mass
\cite{beiner75}, which is increased by the spatial non-locality of the
mean-field potential (Fock terms).


\subsection{The PKO2 versus PKO3 parametrization}

In the framework of shell-model calculations the tensor contribution, arising from pion exchange, has been 
found to play an important role in the description of the evolution of shell structures with 
proton/neutron number \cite{ots05}. The effect of including the pion field in the RHBz model can be 
analyzed using the PKO3 effective interaction \cite{long06a,longthesis}. In the relativistic mean-field framework the pseudo-vector coupling of the pion to the nucleon generates part of the tensor contribution to the effective inter-nucleon interaction, the remaining part being induced by the tensor $\rho$-nucleon coupling. Here we consider the differences between the PKO2 and PKO3 parametrizations on binding energies, ground-state axial deformation parameters, and charge radii. Contrary to the PKO2 effective interaction, the PKO3 effective interaction explicitly includes the pion contribution during the fit to data. The other parameters of the Lagrangian are also affected (Table~\ref{tabpko2}), meaning that the inclusion of the pseudo-vector $\pi-N$ coupling  alters how correlations beyond mean-field are implicitly taken into account through the effective meson couplings.

\begin{figure}[htb]
\begin{center}
\scalebox{0.35}{\includegraphics{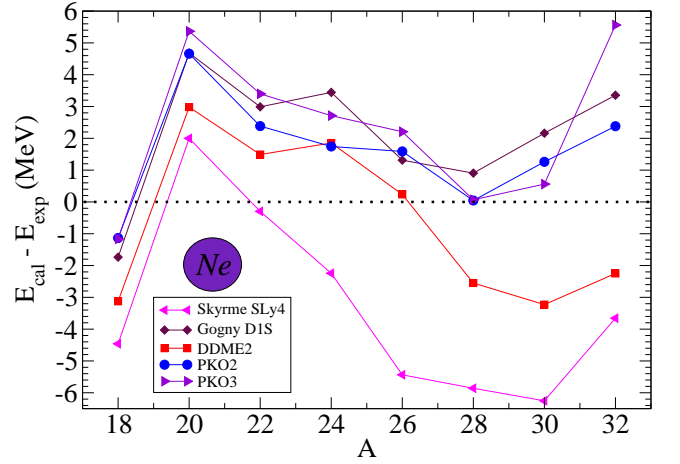}}
\caption{(Color online) Same as in Fig.~\ref{fig:Neediff}, but with the PKO3 RHFBz calculation in addition.}
\label{fig:NeediffPKO3}
\end{center}
 \end{figure}

Figure~\ref{fig:NeediffPKO3} displays the absolute deviations of the theoretical binding energies 
from data for the sequence of neon isotopes. In addition to the results shown in Fig.~\ref{fig:Neediff}, 
here we also include the deviations obtained in the RHFBz calculation with the PKO3 interaction. 
The PKO3 results for the  $\left.\right.^{18}$Ne, $\left.\right.^{26}$Ne,  $\left.\right.^{28}$Ne 
and $\left.\right.^{30}$Ne are on the same 
level of accuracy or better than those obtained with PKO2, i.e.
without the explicit inclusion of the pion field, whereas 
they show a less good agreement with data for the other Ne nuclei. 

\begin{figure}[htb]
\begin{center}
\scalebox{0.35}{\includegraphics{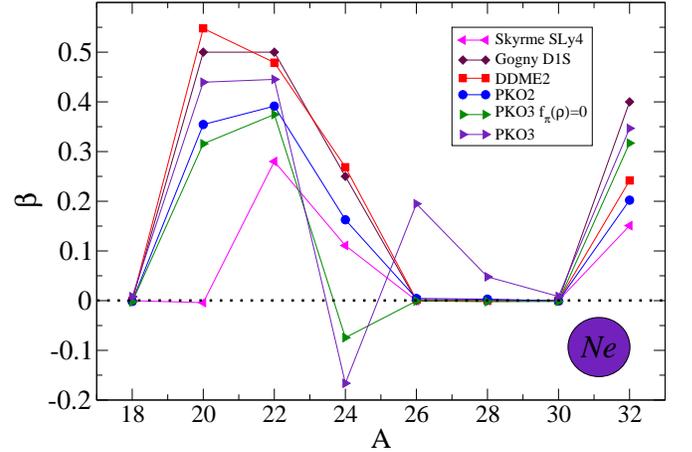}}
\caption{(Color online) Same as in Figure~\ref{fig:Cdef}, but for 
the neon isotopic chain, and with the PKO3 RHFBz calculations and with the PKO3 RHFBz results with $f_\pi(\rho)$ sets to 0 in addition.}
\label{fig:NedefPKO3}
\end{center}
 \end{figure}

The evolution of the axial deformation parameter $\beta$ in the neon isotopic chain is illustrated in Fig.~\ref{fig:NedefPKO3}. In general, the deformation predicted by PKO3 is larger than that calculated with PKO2 and, 
therefore, closer to the results obtained with the DD-ME2 and Gogny D1S effective interactions. PKO3 predicts 
an oblate shape for $\left.\right.^{24}$Ne (quasi degenerate in energy with a prolate solution at $\beta = 0.3$) whereas a prolate ground-state shape for this nucleus is obtained with  PKO2, DDME2, Gogny D1S (quasi degenerate in energy with an oblate solution at $\beta = -0.15$) and Skyrme SLy4 interactions. Moreover, all these interactions, except PKO3 that predicts a prolate ground-state, give no deformation for $\left.\right.^{26}$Ne and $\left.\right.^{28}$Ne. 

\begin{figure}[htb]
\begin{center}
\scalebox{0.35}{\includegraphics{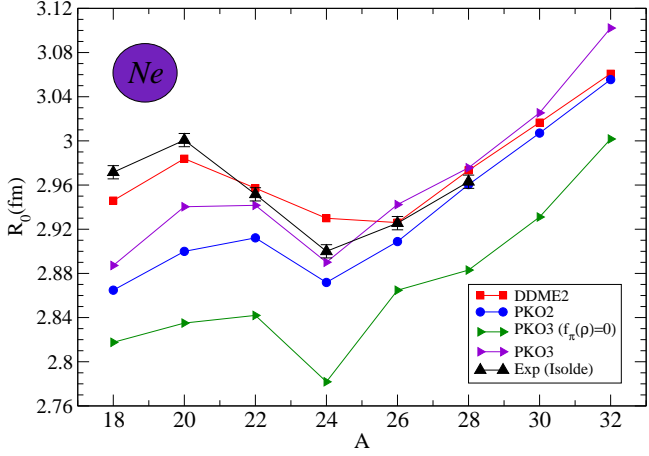}}
\caption{(Color online) Same as in Fig.~\ref{fig:Nerc1}, but with the PKO3 RHFBz calculation in addition.}
\label{fig:Nerc}
\end{center}
 \end{figure}

Finally, in Fig.~\ref{fig:Nerc} we illustrate the differences between the PKO2 and PKO3 parametrizations on the calculated charge 
radii of neon isotopes. The results obtained with PKO3 are shown in comparison with the ISOLDE 
data~\cite{marin}, and with theoretical values predicted by the DD-ME2 (RHB) and PKO2 (RHFBz) 
effective interactions (cf. also Fig.~\ref{fig:Nerc1}). One might notice that 
the explicit inclusion of the 
pion contribution leads to an enhancement of the calculated charge radii as compared with 
the values obtained with PKO2, bringing them closer to the predictions of DD-ME2 and, for 
the lighter isotopes, in better agreement with data. 

\subsection{The effect of including the pion field}

\begin{figure}[htb]
\begin{center}
\scalebox{0.35}{\includegraphics{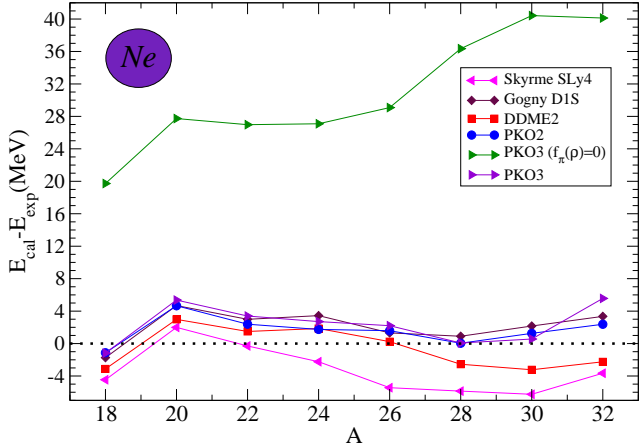}}
\caption{(Color online) Same as in Fig.~\ref{fig:NeediffPKO3}, but with the PKO3 RHFBz calculation with $f_\pi(\rho)=0$ in addition.}
\label{fig:NeediffPKO32}
\end{center}
 \end{figure}

In order to isolate the effect of the pseudo-vector $\pi-N$ coupling
on the observables, we compare calculations based on the PKO3
effective interaction where the pion contribution is switched on and
off. Figure~\ref{fig:NeediffPKO32} displays the absolute deviations of
the theoretical binding energies from data for the sequence of neon
isotopes. In addition to the results shown in
Fig.~\ref{fig:NeediffPKO3}, here we also include the deviations
obtained in the RHFBz calculation with the PKO3 interaction where the
pion contribution is switched off. Switching on the pion coupling
constant brings relevant binding to the Neon isotopes. 

\begin{figure}[htb]
\begin{center}
\scalebox{0.35}{\includegraphics{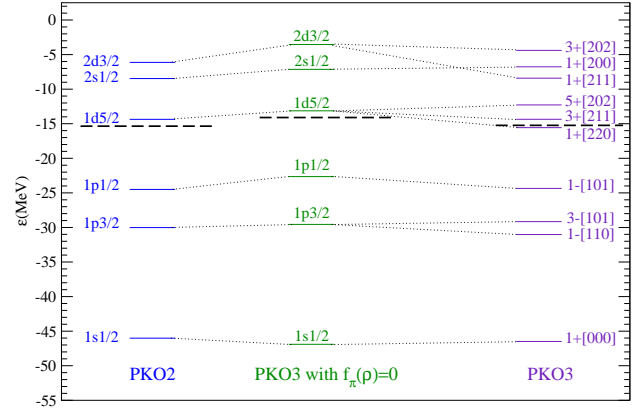}}
\caption{(Color online) Comparison of the single-proton levels of $\left.\right.^{26}$Ne, calculated with the PKO2 effective interaction (on the left), the PKO3 effective interaction where the pion coupling is switched on (on the right) and off (in the middle). The dashed-line denotes the chemical potential.}
\label{fig:Nespp}
\end{center}
\end{figure}

\begin{figure}[htb]
\begin{center}
\scalebox{0.35}{\includegraphics{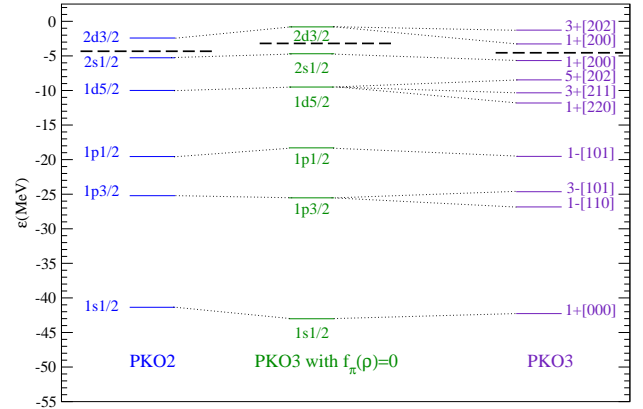}}
\caption{(Color online) Same as in Fig.~\ref{fig:Nespp}, but for the neutron single-particle levels.}
\label{fig:Nespn}
\end{center}
\end{figure}

In Fig.~\ref{fig:NedefPKO3} the comparison between the PKO3 curve and
the PKO3 one where the pion coupling is set to zero shows that the
prolate shape of $^{26,28}$Ne is driven by the pion, whereas
interactions without the explicit tensor term predict a spherical
shape. The effect of the pion field on single-nucleon spectra is
illustrated in Fig.~\ref{fig:Nespp} and~\ref{fig:Nespn}, where we
display the proton and neutron single-particle levels in
$\left.\right.^{26}$Ne obtained in RHFBz calculations with the PKO3
effective interaction where the pion coupling is switched on and off.
PKO3 with the pion coupling set to zero, yields a spherical
ground-state shape and the Nilsson levels are degenerate, in contrast
the degeneracy is lifted in the calculation performed with the
complete PKO3 interaction. Here one notices a clear signature of the
effect of the pion on single-nucleon spectra and the corresponding
evolution of shell structures.

In Fig.~\ref{fig:Nerc} we illustrate the effect of including the pion
field on the calculated charge radii of neon isotopes. Switching on
the pion coupling constant results in a drastic enhancement of
the Neon isotopes charge radii.


\section{Conclusion}
\label{Conclusion}
We have developed the relativistic Hartree-Fock-Bogoliubov model for axially deformed nuclei (RHFBz). 
An effective Lagrangian with density-dependent meson-nucleon couplings is used in the particle-hole channel, and the central part of the Gogny force in the particle-particle channel. The RHFBz quasiparticle equations are solved by expansion in the basis of a deformed harmonic oscillator potential. 
The numerical complexity brought by the explicit treatment of the Fock term within relativistic mean field theory limits, at present, 
 the size of the  oscillator basis for the expansion of the nucleon wave functions. The current version of the model provides a reliable and numerically stable description of ground-state properties up to the zinc isotopic chain. Further numerical optimization is possible and work is in progress to extend the size of the deformed oscillator basis to 12 fermionic shells, allowing a  description of medium-mass and heavy nuclei. In this work illustrative RHFBz calculations have been performed for carbon, neon and magnesium isotopes. Results obtained with the RHF effective force PKO2 have been compared to experimental masses and charge radii and, 
in addition, ground-state deformation and single-nucleon spectra have been shown in comparison with the predictions of one of the most successful RMF meson-exchange interactions: DD-ME2, as well as with the results calculated with the non-relativistic Gogny D1S and Skyrme SLy4 interactions. The effect of explicitly including the pion field has been investigated for binding energies, deformation parameters, and charge radii. The addition of the tensor $\rho$-nucleon coupling will complete the model and thus enable studies of the role of tensor components of the effective inter-nucleon interaction in the evolution of shell structures in deformed nuclei.

\begin{acknowledgements}
The authors acknowledge fruitful discussions with M. Grasso, M. Kowalka, H. Liang, D. Lunney, 
T. Nik$\breve{s}$i\'c, P. Ring, N. Van Giai and D. Verney. This
work was supported in part by the ANR Nexen and MZOS - project 1191005-1010.
\end{acknowledgements}

\appendix
\section{Explicit expressions for the exchange contribution}
\label{app}

Each matrix $A$, $B$, and $C$ in Eq.~(\ref{eq:mat1}) contains the kinetic, direct, exchange and rearrangement contributions. The explicit expressions for the kinetic and direct contributions read:
\begin{eqnarray}
&&\left( \begin{array} {c}
 A_{\alpha,\alpha'} \\
 C_{\alpha,\alpha'}  \end{array} \right)  = \delta_{m_l,m_l'} \delta_{m_s,m_s'} N_{n_r}^{ml} N_{n_z} N_{n_r'}^{ml'} N_{n_z'} \nonumber \\ && \int_0^\infty d^\eta e^{-\eta} \eta^{m_l} L_{n_r}^{ml}(\eta) L_{n_r'}^{ml'}(\eta) \int_0^\infty d^\zeta e^{-\zeta^2} H_{n_z}(\zeta) H_{n_z'}(\zeta) \nonumber \\ && (M^*(b\perp \sqrt{\eta},b_z \zeta)\pm V(b\perp \sqrt{\eta},b_z \zeta)) \nonumber \\
\end{eqnarray}
\begin{eqnarray}
&&B_{\alpha,\alpha'} = \delta_{m_l,m_l'} \delta_{m_s,m_s'} \delta_{n_r,n_r'} \frac{(-1)^{-m_s+\frac{1}{2}}}{b_z} \nonumber \\ &&\left( \delta_{n_z',n_z+1}\sqrt{\frac{n_z'}{2}} -\delta_{n_z,n_z'+1}\sqrt{\frac{n_z}{2}} \right)  + \delta_{m_l,m_l'} \delta_{n_z,n_z'} \frac{N_{n_r}^{ml}N_{n_r'}^{ml'}}{b_\perp} \nonumber \\
&& \times \left\lbrace \delta_{m_s',m_s+1}\int_0^\infty d^\eta e^{-\eta} \eta^{m_l-\frac{1}{2}} L_{n_r}^{ml}(\eta)\right.  \nonumber \\ &&  \left. \left[\tilde{L}_{n_r'}^{ml}(\eta)+(1-m_l)L_{n_r'}^{ml}(\eta) \right] \right.   \nonumber \\
&& +\delta_{m_s,m_s'+1}\int_0^\infty d^\eta e^{-\eta} \eta^{m_l-\frac{1}{2}} L_{n_r}^{ml}(\eta) \nonumber \\ && \left. \left[\tilde{L}_{n_r'}^{ml}(\eta)+(1+m_l)L_{n_r'}^{ml}(\eta) \right] \right\rbrace   \nonumber \\ 
\end{eqnarray}
where
\begin{eqnarray}
&& \tilde{L}_{n_r}^{ml}(\eta) = \left(2m_l+n_r-\eta \right) L_{n_r}^{ml}(\eta) - 2 \left(n_r+m_l \right)L_{n_r-1}^{ml}(\eta) \nonumber \\
\end{eqnarray}
Taking the $\sigma$ meson as example, the exchange contribution to the $A$, $B$ and $C$ matrices is given by the following expressions:
\begin{eqnarray}
A_{\alpha,\alpha'}^{\sigma}(q_i) &=& \sum_{j>0}\delta_{q_j,q_i} \sum_{\beta,\beta'}f_{\beta}^{(j)}(q_j) f_{\beta'}^{(j)}(q_j)\times \nonumber \\
&&   \left\lbrace \delta^\shortparallel_{m_{s_{\beta'\alpha'}}} \delta^\shortparallel_{m_{s_{\beta \alpha}}} I^\sigma_{\alpha\beta\beta'\alpha'}\right.  + \nonumber \\  &&\left. \delta^\nshortparallel_{m_{s_{\beta'\alpha'}},0}\delta^\nshortparallel_{m_{s_{\beta\alpha}}}(2.ms_\beta)(2.ms_{\beta'}) \breve{I}^\sigma_{\alpha\beta\beta'\alpha'}\right\rbrace \nonumber \\
\end{eqnarray}

\begin{eqnarray}
B_{\alpha,\tilde{\alpha'}}^{\sigma}(q_i) = && -\sum_{j>0}\delta_{q_j,q_i} \sum_{\beta,\tilde{\beta'}} f_{\beta}^{(j)}(q_j) g_{\tilde{\beta'}}^{(j)}(q_j)\times \nonumber \\ &&  \left\lbrace \delta^\shortparallel_{m_{s_{\tilde{\beta'}\tilde{\alpha'}}}}\delta^\shortparallel_{m_{s_{\beta\alpha}}} I^\sigma_{\alpha\beta\tilde{\beta'}\tilde{\alpha'}}\right.  - \nonumber \\ && \left. \delta^\nshortparallel_{m_{s_{\tilde{\beta'}\tilde{\alpha'}}}}\delta^\nshortparallel_{m_{s_{\beta\alpha}}}(2.ms_\beta)(2.ms_{\tilde{\beta'}}) \breve{I}^\sigma_{\alpha\beta\tilde{\beta'}\tilde{\alpha'}}  \right\rbrace \nonumber \\
\end{eqnarray}

\begin{eqnarray}
C_{\tilde{\alpha},\tilde{\alpha'}}^{\sigma}(q_i) = && \sum_{j>0}\delta_{q_j,q_i} \sum_{\tilde{\beta},\tilde{\beta'}} g_{\tilde{\beta}}^{(j)}(q_j) g_{\tilde{\beta'}}^{(j)}(q_j) \times\nonumber \\
&&  \left\lbrace  \delta^\shortparallel_{m_{s_{\tilde{\beta'}\tilde{\alpha'}}}}\delta^\shortparallel_{m_{s_{\tilde{\beta}\tilde{\alpha}}}} I^\sigma_{\tilde{\alpha}\tilde{\beta}\tilde{\beta'}\tilde{\alpha'}} \right.  +  \nonumber \\ &&\left. \delta^\nshortparallel_{m_{s_{\tilde{\beta'}\tilde{\alpha'}}}}\delta^\nshortparallel_{m_{s_{\tilde{\beta}\tilde{\alpha}}}}(2.ms_{\tilde{\beta}})(2.ms_{\tilde{\beta'}}) \breve{I}^\sigma_{\tilde{\alpha}\tilde{\beta}\tilde{\beta'}\tilde{\alpha'}}     \right\rbrace \nonumber \\
\end{eqnarray}
where
\begin{eqnarray}
&& \delta^\shortparallel_{m_{s_{\beta\alpha}}} \equiv \delta_{m_{s_{\beta}},m_{s_{\alpha}}} \\
&&\delta^\nshortparallel_{m_{s_{\beta\alpha}}} \equiv \delta_{m_{s_{\beta}},-m_{s_{\alpha}}} 
\end{eqnarray}
\begin{eqnarray}\label{fockint}
&& I^\sigma_{\alpha\beta\beta'\alpha'}\equiv\int{d{\bf r}[g_\sigma\phi_\alpha^*\phi_\beta](\bm{r})} \int {d{\bf
r'} D_\sigma(\bm{r},\bm{r'})[g_\sigma\phi_{\beta'}^*\phi_{\alpha'}](\bm{r'})} \nonumber \\
\\
&& \breve{I}^\sigma_{\alpha\beta\beta'\alpha'} \equiv \int{d{\bf r}[g_\sigma\phi_\alpha^*\phi_\beta^*](\bm{r})} \int {d{\bf
r'} D_\sigma(\bm{r},\bm{r'})[g_\sigma\phi_{\beta'}\phi_{\alpha'}](\bm{r'})} \nonumber \\
\end{eqnarray}
The Fock terms are characterized by non-degenerate contributions to the positive and negative $\Omega_j$ blocks. The integrals can be written as:
 \begin{eqnarray}
&& I^m_{\alpha\beta\beta'\alpha'} = (-1)^{(m_{l_{\alpha'}}-m_{l_{\beta'}})} \delta_{(m_{l_\beta}-m_{l_\alpha})+(m_{l_{\alpha'}}-m_{l_{\beta'}}),0} 
\nonumber \\ && \phantom{aaa} \int {\frac{d{k_\perp}d{k_z}}{(2\pi)^2} Q_{\alpha,\beta}(k_\perp, k_z) \frac{k_\perp}{k_\perp^2+k_z^2+m_m^2} Q_{\beta',\alpha'}(k_\perp,-k_z)} \nonumber \\
\end{eqnarray}
where
\begin{eqnarray} \label{fncQ}
&& Q_{\alpha,\beta}(k_\perp,k_z)=\int d{r_{1\perp}}d{z_1}r_{1\perp}[g_\sigma \check{\phi_\alpha}\check{\phi_\beta}](r_{1\perp},z_1) \nonumber \\ && \phantom{aaaaaaaa} e^{ik_z z_1} J_{(m_{l_\beta}-m_{l_\alpha})}(k_\perp r_{1\perp})    \nonumber \\
&& Q_{\beta',\alpha'}(k_\perp,-k_z)=\int d{r_{2\perp}}d{z_2}r_{2\perp}[g_\sigma \check{\phi_{\beta'}}\check{\phi_{\alpha'}}](r_{2\perp},z_2)\nonumber \\ && \phantom{aaaaaaaa} e^{-ik_z z_2} J_{(m_{l_{\alpha'}}-m_{l_{\beta'}})}(k_\perp r_{2\perp})
\end{eqnarray}

Taking again the case of the $\sigma$ meson field as example, the exchange contribution to the rearrangement term reads:
\bea \label{rearex}
\Sigma^{\sigma,Ex}_R(\bm{r}) = && \sum_{m,n}\delta_{q_m,q_n} [\frac{\partial g_\sigma}{\partial \rho_v}\bar{f_m}(q_m)f_n(q_n) ](\bm{r}) \nonumber \\ && \int d^3r' \left\lbrace D_\sigma(\bm{r},\bm{r'})[g_\sigma \bar{f_n}(q_n)f_m(q_m)](\bm{r'}) \right\rbrace \nonumber \\ 
\eea
In the basis of a deformed oscillator, relation (\ref{rearex}) takes the form :
\bea
\Sigma_{R;\alpha\alpha'}^{\sigma,Ex} = && \sum_{m>0}\sum_{\mu\mu'} f_\mu^{(m)}(q_m)f_{\mu'}^{(m)}(q_m) \tilde{A}_{\alpha\alpha'\mu\mu'}^\sigma \nonumber \\ &&+ 2 \sum_{m>0}\sum_{\tilde{\mu}\mu'} g_{\tilde{\mu}}^{(m)}(q_m)f_{\mu'}^{(m)}(q_m) \tilde{B}_{\alpha\alpha'\tilde{\mu}\mu'}^\sigma \nonumber \\
&& + \sum_{m>0}\sum_{\tilde{\mu}\tilde{\mu'}} g_{\tilde{\mu}}^{(m)}(q_m)g_{\tilde{\mu'}}^{(m)}(q_m) \tilde{C}_{\alpha\alpha'\tilde{\mu}\tilde{\mu'}}^\sigma
\eea
The matrices $\tilde{A}$,  $\tilde{B}$ and $\tilde{C}$ are obtained by replacing the integrals $I^\sigma_{\mu\nu\nu'\mu'}$ and $\breve{I}^\sigma_{\mu\nu\nu'\mu'}$ in the expressions for 
the corresponding matrices $A$, $B$ and $C$, with the integrals $K^\sigma_{\alpha\alpha'\mu\nu;\nu'\mu'}$ and $\breve{K}^\sigma_{\alpha\alpha'\mu\nu;\nu'\mu'}$:
\bea
K_{\alpha,\alpha',\gamma,\lambda,\lambda',\gamma'}^m= &&\int{d{\bf r}[\phi_\alpha^*\phi_{\alpha'}\frac{\partial g_m}{\partial \rho_v}\phi_\gamma^*\phi_\lambda ](\bm{r})} \nonumber \\ && \int {d{\bf
r'} D_m(\bm{r},\bm{r'})[g_m\phi_{\lambda'}^*\phi_{\gamma'}](\bm{r'})}
\eea
\bea
\breve{K}_{\alpha,\alpha',\gamma,\lambda,\lambda',\gamma'}^m= &&\int{d{\bf r}[\phi_\alpha^*\phi_{\alpha'}\frac{\partial g_m}{\partial \rho_v}\phi_\gamma^*\phi_\lambda^* ](\bm{r})} \nonumber \\ && \int {d{\bf
r'} D_m(\bm{r},\bm{r'})[g_m\phi_{\lambda'}\phi_{\gamma'}](\bm{r'})}
\eea
These integrals can be written in the form:
\bea
&& K_{\alpha,\alpha',\gamma,\lambda,\lambda',\gamma'} =  \delta_{(m_{l_{\alpha'}}-m_{l_\alpha}+m_{l_\lambda}-m_{l_\gamma})+(m_{l_{\gamma'}}-m_{l_{\lambda'}}),0} \nonumber \\
&& \int \frac{d{k_\perp}d{k_z}}{(2\pi)^2} Q_{\alpha,\alpha',\gamma,\lambda}(k_\perp, k_z) \frac{k_\perp}{k_\perp^2+k_z^2+m_m^2} Q_{\lambda',\gamma'}(-k_\perp,-k_z) \nonumber \\ 
\eea
where
\begin{eqnarray}
&& Q_{\alpha,\alpha',\gamma,\lambda}(k_\perp,k_z)=\int d{r_{1\perp}}d{z_1}r_{1\perp} [ \check{\phi_\alpha}\check{\phi_{\alpha'}}\frac{\partial g_m }{\partial \rho_v} \check{\phi_\gamma}\check{\phi_\lambda}](r_{1\perp},z_1) \nonumber \\ && e^{ik_z z_1} J_{(m_{l_{\alpha'}}-m_{l_\alpha}+m_{l_\lambda}-m_{l_\gamma})}(k_\perp r_{1\perp}) 
\\
&& Q_{\lambda',\gamma'}(-k_\perp,-k_z)=\int d{r_{2\perp}}d{z_2}r_{2\perp}[g_m \check{\phi_{\lambda'}}\check{\phi_{\gamma'}}](r_{2\perp},z_2) \nonumber \\ && e^{-ik_z z_2} J_{(m_{l_{\gamma'}}-m_{l_{\lambda'}})}(-k_\perp r_{2\perp})   
\end{eqnarray}
and

\bea
&&\breve{K}_{\alpha,\alpha',\gamma,\lambda,\lambda',\gamma'} =  \delta_{(m_{l_{\alpha'}}-m_{l_\alpha}-m_{l_\lambda}-m_{l_\gamma})+(m_{l_{\gamma'}}+m_{l_{\lambda'}}),0} \nonumber \\
&& \int {\frac{d{k_\perp}d{k_z}}{(2\pi)^2} Q_{\alpha,\alpha',\gamma,\lambda}^-(k_\perp, k_z) \frac{k_\perp}{k_\perp^2+k_z^2+m_m^2} Q_{\lambda',\gamma'}^+(-k_\perp,-k_z)} \nonumber \\
\eea
where
\begin{eqnarray}
&& Q_{\alpha,\alpha',\gamma,\lambda}^-(k_\perp,k_z)=\int d{r_{1\perp}}d{z_1}r_{1\perp}[ \check{\phi_\alpha}\check{\phi_{\alpha'}}\frac{\partial g_m }{\partial \rho_v} \check{\phi_\gamma}\check{\phi_\lambda}](r_{1\perp},z_1) \nonumber \\ && e^{ik_z z_1} J_{(m_{l_{\alpha'}}-m_{l_\alpha}-m_{l_\lambda}-m_{l_\gamma})}(k_\perp r_{1\perp})   
\\
&& Q_{\lambda',\gamma'}^+(-k_\perp,-k_z)=\int d{r_{2\perp}}d{z_2}r_{2\perp}[g_m \check{\phi_{\lambda'}}\check{\phi_{\gamma'}}](r_{2\perp},z_2) \nonumber \\ && e^{-ik_z z_2} J_{(m_{l_{\gamma'}}+m_{l_{\lambda'}})}(-k_\perp r_{2\perp})   
\end{eqnarray}


\begin{thebibliography}{99}

\bibitem{lac09} D. Lacroix, T. Duguet and M. Bender, Phys. Rev. \textbf{C79}, 044318 (2009).

\bibitem{bend03} M. Bender, P.-H. Heenen and P.-G. Reinhard, Rev. Mod. Phys. \textbf{75}, 121 (2003).

\bibitem{vret05} D. Vretenar, A.V. Afanasjev, G.A. Lalazissis and P. Ring, Phys. Rep. \textbf{409}, 101 (2005).

\bibitem{FS.00} R. J. Furnstahl and B. D. Serot, Comments Nucl. Part. Phys. 2, A23 (2000).

\bibitem{AA10} A. V. Afanasjev and H. Abusara, Phys. Rev. \textbf{C81}, 014309 (2010)  
	
\bibitem{Joe.05} J. N. Ginocchio, Phys. Rep. 414, 165 (2005).

\bibitem{bouy87} A. Bouyssy, J.-F. Mathiot, N. Van Giai and S. Marcos, Phys. Rev. \textbf{C36}, 380 (1987).

\bibitem{bern93} P. Bernardos, V.N. Fomenko, N. Van Giai, M.L. Quelle, S. Marcos, R. Niembro, L.N. Savushkin, Phys. Rev. \textbf{C48}, 2665 (1993).

\bibitem{long06} W.H. Long, N. Van Giai and J. Meng, Phys. Lett. \textbf{B640}, 150 (2006).

\bibitem{long10} W.H. Long, P. Ring, N. Van Giai and J. Meng, Phys. Rev. \textbf{C81}, 024308 (2010).





\bibitem{long08} W.H. Long, H. Sagawa, J. Meng and N. Van Giai, Europhysics Letters \textbf{82}, 12001 (2008).

\bibitem{long07} W.H. Long, H. Sagawa, N. Van Giai and J. Meng, Phys. Rev. \textbf{C76}, 034314 (2007).

\bibitem{liang08} H. Liang, N. Van Giai and J. Meng, Phys. Rev. Lett. \textbf{101}, 122502 (2008). 

\bibitem{liang10} H. Liang, W.H. Long, J. Meng and N. Van Giai, Eur. Phys. J. \textbf{A44},119-124 (2010).

\bibitem{ots05} T. Otsuka, T. Suzuki, R. Fujimoto, H. Grawe and Y. Akaishi, Phys. Rev. Lett. \textbf{95}, 232502 (2005).

\bibitem{paar04} N. Paar, T. Nik$\breve{s}$i\'c, D. Vretenar and P. Ring, Phys. Rev. \textbf{C69}, 054303 (2004).

\bibitem{NVR.11} T. Nik{\v{s}}i{\'{c}}, D. Vretenar, and P. Ring, Prog. Part. Nucl. Phys. \textbf{66}, 519-548 (2011).

\bibitem{long06a} W.H. Long, N. Van Giai and J. Meng, arXiv:nucl-th/0608009.

\bibitem{longthesis} W.H. Long, PhD thesis, Universite Paris-Sud 11 (2005).

\bibitem{NVF.02} T. Nik{\v{s}}i{\'{c}}, D. Vretenar, P. Finelli, and P. Ring, Phys. Rev. {\bf C 66},  024306 (2002).

\bibitem{TW.99} S. Typel and H.~H. Wolter, Nucl. Phys. {\bf A656},  331  (1999).

\bibitem{furn04} R.J. Furnstahl, Lecture Notes in Physics \textbf{641}, 1-29 (2004).

\bibitem{ring97} P. Ring, Y.K. Gambhir and G.A. Lalazissis, Comput. Phys. Commun. \textbf{105}, 77-97 (1997).

\bibitem{rs80} P. Ring, P. Schuck, {\it The nuclear many-body problem,
Springer-Verlag} (1980).

\bibitem {BGG.84}J.~F. Berger, M. Girod, and D. Gogny, Nucl. Phys. A 428, 23c (1984).

\bibitem {BGG.91}J.~F. Berger, M. Girod, and D. Gogny, Comp. Phys. Comm. 63, 365 (1991).

\bibitem{webGogny} S. Hilaire and M. Girod, Eur. Phys. J. \textbf{A33},237-241 (2007).

\bibitem {lal05}G.A. Lalazissis, T. Nik\v{s}i\'{c}, D. Vretenar and P. Ring,
Phys. Rev. C 71, 024312 (2005).

\bibitem{webSkyrme} M. V. Stoitsov, J. Dobaczewski, W. Nazarewicz, S. Pittel, and D. J. Dean, Phys. Rev. \textbf{C68}, 054312 (2003) .

\bibitem{audi93} G. Audi and W.H. Wapstra, Nucl. Phys. \textbf{A565},1 (1993).

\bibitem{sag04} H. Sagawa, X.R. Zhou, X.Z. Zhang and T. Suzuki, Phys. Rev.\textbf{C70}, 054316 (2004).

\bibitem{ima04} N. Imai et. al., Phys. Rev. Lett. \textbf{92}, 062501 (2004).

\bibitem{ong06} H.J. Ong et. al., Phys. Rev.\textbf{C73}, 024610 (2006).

\bibitem{hag07} K. Hagino and H. Sagawa, Phys. Rev.\textbf{C75}, 021301 (2007).

\bibitem{wied08} M. Wiedeking \textit{et al.}, Phys. Rev. Lett. \textbf{100}, 152501 (2008).

\bibitem{wuo10} A.H. Wuosmaa \textit{et al.}, Phys. Rev. Lett. \textbf{105}, 132501 (2010).

\bibitem{marin} K. Marinova \textit{et al.}, to be submitted.

\bibitem{beiner75} M. Beiner \textit{et al}, Nucl. Phys. \textbf{A238}, 29 (1975).


\end{thebibliography}
\end{document}